# Numerical Realization of Bethe Rapidities in cold quenched systems- a search for scale invariance

Sumita Datta, Department of Physics, The Icfai Foundation for Higher Education , Hyderabad-501203, India

Maxim Olshanii, Department of Physics, University of Massachusetts, Boston MA 02125, USA

We numerically investigate on the scale invariance of non-equilibrium problem associated with the (i) release of a cold Lieb-Liniger[1] Bose gas from a harmonic trap and (ii) a ballistic expansion of a LL gas from the ground state of an infinitely deep hard wall box of length $L_0$ to L . Now in asymptotically long time-of-flight experiments, the momentum distribution coincides with the shape of the real space particle density. We exploit this picture to analyze the scale invariance in the above situations and find that only for release from a harmonic trap and in the limit of weak and strong interaction one can recover self similar distribution whereas for the free expansion of a LL gas from an infinitely deep box, the density neither follows the self similar evolution and nor scale invariance. These numerical findings are in good agreement with the pioneering work of Campbell et al[2] and del Campo et al[3]

**Introduction**: Now only in the asymptotically long time-of-flight experiments, one can use the Bethe rapidities[4] to describe the long time density profile. But this requires an assumption of ballistic dynamics and is an approximation which becomes good only at long times of evolution. In this work we study a system of strongly interacting LL bosons in a harmonic trap and watch for its density profile right after the (i) release of LL bosons from a harmonic trap and (ii) during its ballistic expansion from the ground state in an infinitely deep box of length $L_0$ to L. There are several motivations for considering this. Firstly we can have a much higher level control over such systems. These few body systems are amenable to numerical techniques which help us understanding the different features of the system without resort to uncontrolled approximations. The interesting aspects of a strongly correlated system is that it is experimentally accessible only for small systems and quantum effects become pronounced in such systems of reduced dimensionality. These 1d systems are extremely interesting and worth investigating in the context of non-equilibrium dynamics of interacting many body systems. But the most interesting thing is we want to suggest an experimental procedure that would allow for a direct observation of distribution of Bethe rapidities. This rapidities play a key role in deciding the scale invariance and can work as a 'workhorse' in the theory of 'Short cut to adiabaticity[5]

We consider two quench protocols: (i) We consider a LL gas in a harmonic trap and watch for its density evolution for weak and strong interactions by releasing it from the trap.(ii) We study the density evolution of a LL gas during its expansion from the ground state in an infinitely deep box of length $L_0$ to L. We want to expand the box to a large dimension, large enough so that most particles do not have the time to reach its wall by the end of the expansion. After an expansion to a larger cloud size much larger than $L_0$ the spatial density distribution will be proportional to the density of Bethe rapidities.

We choose a Bose gas in which the atomic wave guides have transverse degrees of freedom frozen making it effectively an 1D system. These systems can be described in terms of Lieb-Liniger model in which the identical bosons interact with a delta function interaction of arbitrary strength 'c'. In the limit of $c \to \infty$, the Lieb-Liniger gas approaches the Tonks-Girardeau[6] regime of impenetrable bosons. In this limit finding the time dependent solutions become simpler due to Bose-Fermi mapping. In the case of finite interaction strength c it is much more difficult to calculate exact may body wave-functions and other observables describing the dynamics of time-dependent Lieb-Liniger wave packets. We apply a quantum Monte Carlo method based on Feynman-Kac[7] method to calculate the density profile of bose gas with finite interaction during the expansion of the box without any difficulty. As a matter of fact we solve the time dependent Schrödinger equation for the system of one-dimensional bosons interacting via delta potential in an infinite square well (namely Lieb-Liniger model) using Feynman-Kac path integral Monte Carlo technique. Even though the systems can be realized experimentally and are exactly solvable by Bethe Ansatz[8,9], the diffusion Monte Carlo is proven to be more efficient in most circumstances than other mean value techniques as the numerical method can incorporate the finite interaction very easily. Using N particle ground state wavefunction for one–dimensional hard core bosons in a harmonic trap, we develop an algorithm to calculate the density. We also observe the change in the density distribution of the LL gas after its release from a harmonic trap and with the increase in the length of the hard wall box. Bethe ansatz has been previously used to calculate the eigenstates of LL bosons on infinite line[1] with periodic boundary condition and in an infinitely deep box [10,11].

In section 1 we describe the theory, in section 2 we show the numerical results and in section 3 we have discussion and future plans.

# 1 The model and the quench protocol:

## 1.1 Theory

We numerically study the free expansion of a density of a 1 dimensional system (namely Lieb Liniger (LL) gas) of N $Rb^{87}$ atoms which was initially in the ground state of an infinitely deep hard wall potential by Feynman-Kac path integral technique. Let us consider a Bose-Einstein condensate of N ($N \gg 1$) atoms of mass m trapped in 1 dimensional harmonic potential $V_{trap}$ and interacting with a delta interaction $V_{int}$. The LL bose gas at low density and temperature is studied in the Tonks-Girardeau limit as the interaction strength is finite but very large. In the Schrödinger picture the Hamiltonian then can be represented as

$$H = -\Delta/2 + V \qquad (1)$$

where $V = V_{trap} + V_{int}$

$$\frac{\hbar^2}{2m}\sum_{i=1}^{N}(-\frac{\partial^2 \psi}{\partial y_i^2} + \frac{1}{2}m\omega^2 y_i^2) + g_{1D}\sum_{i<j}\delta(y_i - y_j)\psi = \tilde{E}\psi$$

$$D: -\infty \langle y_1 \leq y_2 \leq ...... \leq y_N \langle \infty$$

with $\hbar = m = 1$, one gets

$$\sum_{i=1}^{N}(-\frac{1}{2}\frac{\partial^2 \psi}{\partial y_i^2} + \frac{1}{2}\frac{\omega^2 y_i^2}{\hbar^2}) + V\psi = \frac{E}{\hbar^2}\psi$$

$V(y) = g_{1D}\delta_\sigma(y)$ where $g_{1D}$ is the effective one dimensional coupling strength[12]

**Case I: 1D delta interacting bosons in an Harmonic trap on an infinite line.**

The purpose of these simulations was to watch the variation in density profile with (i) release of LL gas from a harmonic trap and (ii) a ballistic expansion from the ground state of box of length $L_0$ to L. The idea is after an expansion to a cloud size much greater $L_0$, the spatial distribution will be proportional to the density of Bethe rapidities. We are interested in the Bethe rapidities as we want examine the self similarity and scale invariance associated with the above non-equilibrium phenomena.

We are now interested in the quasi-1D case where the trap is supposed to be anisotropic such that transverse characteristic length $a_\perp \langle\langle a_\parallel$. We will work with a Hamiltonian rescaled[13] to the length scale of 1D longitudinal system, $a_\parallel$. Putting $y_i = y_i^{'} a_\parallel$

$$\sum_{i=1}^{N}(-\frac{1}{2}\frac{\partial^2 \psi}{\partial y_i^{'2}} + \frac{1}{2}y_i^{'2}\psi + V^{'}(y_i^{'}))\psi = E^{'}\psi$$

$$D: -\infty \langle y_1^{'} \leq y_2^{'} \leq ...... \leq y_N^{'} \langle \infty$$

$V^{'}(y_i^{'}) = g_{1D}^{'}\delta_\sigma(y_i^{'})$, $g_{1D}^{'} = \frac{2a_0^{'}}{a_\perp^{'2}}(1 - C\frac{a_0^{'}}{a_\perp^{'}})^{-1}$; $a_0^{'} = a_0/a_\parallel$ and $a_\perp^{'} = a\perp/a_\parallel$

**Case II: Ballistic Expansion for 1D bosons delta interacting bosons in a hard wall trap from a box length of $L_0$ to L**

Next we set the ground state of N delta interacting bosons( $Rb^{87}$ ) in an infinite square well. The gas is then released from a harmonic trap and allowed to expand [14,15] with the interaction on from a box of length L=150 nm to L=600 nm which pictorially can be represented as follows:

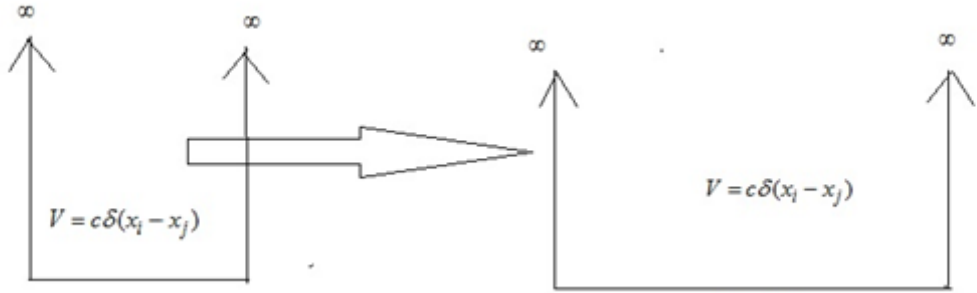

Fig 2 Ballistic exapnsion of the interacting trapped bose gas from a box leghth L=150 to L=600.

The stationary Schrodinger equation for N equal mass particles( $Rb^{87}$ ) in 1-dimension with two body delta function interaction reads as

$$\sum_{i=1}^{N}(-\frac{1}{2}\frac{\partial^2 \psi}{\partial y_i^{'2}} + V'(y_i')\psi + \frac{1}{2}\frac{\omega^2 y_i^{'2}}{\hbar^2}\psi) = \tilde{E}'\psi \quad ; \tilde{c} > 0 \tag{10}$$

$D: 0 < y_1 \leq y_2 \leq \ldots \leq y_N < L$

$V'(y_i') = g_{1D}' \delta_\sigma(y_i') \; ; \; g_{1D}' = \frac{2a_0'}{a_\perp^{'2}}(1 - C\frac{a_0'}{a_\perp'})^{-1}$

$\delta_\sigma(y_i') = \frac{\exp\left(-\frac{y_i^{'2}}{2\sigma^2}\right)}{\sqrt{2\pi}\sigma}$ which tends to $\delta$ as $\sigma \to 0$ in the distribution sense.

$$\sum_{i=1}^{N}(-\frac{1}{2}\frac{\partial^2 \psi}{\partial y_i^{'2}} + \frac{g_{1D}' \exp\left(-\frac{y_i^{'2}}{2\sigma^2}\right)}{\sqrt{2\pi}\sigma}\psi + \frac{1}{2}y_i^{'2}\psi) = \tilde{E}\psi$$

Now if we now consider the box of length 'L' instead of indefinite axis it is better to rescale[16] the length scale in terms of 'L' as follows:

Now rescaling as $x_i = y_i'/L$, $E = \tilde{E}L^2$ and defining $c = \frac{g_{1D}'L}{\sqrt{2\pi}\sigma}$ , we get

$$\sum_{i=1}^{N}(-\frac{1}{2}\frac{\partial^2 \psi}{\partial x_i^2} + c\exp\left(-\frac{x_i^2}{2\sigma^2}\right)\psi + \frac{1}{2}x_i^2\psi) = E\psi$$

## 1.2. Path integral formalism in Quantum Mechanics as a basis for the numerical simulations.

Mostly at T=0, Gross-Pitaevski (GP) technique is applied for the calculation of energy, density of a many body system. The mean field approach such as GP for solving many body dynamics is only approximate. We propose to apply Diffusion Monte Carlo technique known as Feynman-Kac path integral simulation to the above quantum gas. To connect Feynman-Kac ( FK ) or Generalized Feynman-Kac ( GFK) to other many body techniques our numerical procedure has a straight-forward implementation[17] to Schrödinger's wave mechanics.

Let us consider the time-dependent Schrödinger equation with a single particle with Hamiltonian

$$H = (-\frac{\Delta}{2} + V) \text{ i.e., }$$

$$i\frac{\partial u(x,t)}{\partial t} = (-\frac{\Delta}{2} + V) u(x,t) \tag{1}$$

u(x,0)=f(x)

Multiplication of the above equation by –i leads to the evolution equation

$$\left.\begin{array}{l}\frac{\partial u}{\partial t} = -iHu \\ u(0) = f\end{array}\right\}$$

whose solution has the form

$$u(x,t) = \left(e^{-itH} f\right)(x)$$

To interpret to Feynman path integral we use Trotter's product formula stated in the following theorem :

**Theorem** Let A, B and A+B be self-adjoint opertors on a Hilbert space, then

$$e^{-it(A+B)} = \lim_{n \to \infty} (e^{-itA/n} e^{-itB/n})^n$$

Applying Trotter product formula one can write

$$e^{-itH} f(x_0) = \lim_{n \to \infty} (e^{-(\frac{it}{n})H_0} e^{-(\frac{it}{n})V})^n f(x_0)$$

$$= \lim_{n \to \infty} \int_{R^{3n}} e^{iS_n(x_0, x_1, \ldots, t)} f(x_n) dx_n \ldots dx_1 \text{ where } S_n(x_0, x_1, \ldots, t) = \sum_{i=1}^{n} \frac{t}{n} \left[\frac{1}{2} \frac{|x_i - x_{i-1}|^2}{t/n} - V(x_i)\right]$$

$$= \lim_{n \to \infty} \int_{R^{3n}} e^{iS_n(x_0, x_1, \ldots, t)} f(x_n) d\mu_n$$

$$= \int_{R^{\infty}} e^{iS(x_0, x_1, \ldots, t)} f(x) d\mu \text{ provided } d\mu_n = (\frac{2\pi it}{n})^{-3n/2} dx_n \ldots dx_1 \text{ converges to } d\mu \text{ and } S_n \text{ converges to } S \text{ as } n \to \infty$$

Unfortunately, one cannot justify the existence of S and $\mu$

because of the following reasons:

(a) most paths are not differentiable as a function of s

(b) $d\mu_n = (\frac{2\pi it}{n})^{-3n/2} dx_n.....dx_1$ does not exist as $n \to \infty$.

To overcome the difficulty associated with the non-existence of complex valued $\mu(x)$ on $R^\infty$ described above, one replaces t by –it in equation(1) and obtains a diffusion equation :

$$\frac{\partial u(x,t)}{\partial t} = (\frac{\Delta}{2} - V)u(x,t)$$

$$u(x,0) = f(x)$$

$$u(x,t) = (e^{-tH}f)(x) = \int_\Omega \prod_{j=1}^{n} e^{-\frac{t}{n}V(x_j)} f d\tilde{\mu}_{x_0}^n$$

where unlike in the previously considered complex case

$d\tilde{\mu}_{x_0}^n = dx_1........dx_n k(x_0, x_1; t/n).........k(x_{n-1}, x_n; t/n)$ with $k(x, y, t) = (2\pi t)^{-n/2} e^{-(x-y)^2/2t}$, is a real valued measure which has a limit as $n \to \infty$. Namely $\tilde{\mu}_{x_0}^n$ can be identified with a probability measure i.e., Wiener measure. In terms of the propagator the solution of eqn(1) can be written as

$$u(x,t) = \int k(x, y, t) u(y) dy$$

This really shows thta the above solution is the time dependent as propagator $k(x, y, t)$
Is time dependent.

It shows that Feynman-Kac solution can handle non-stationary problems like the one we are considering in this paper.

The above solution can be given a probabilistic flavor by noticing

$$u(x,t) = E_x \left\{ e^{-\int_0^t V(x(s))ds} f(X(t)) \right\}$$ where $E_x$ is the expected value of the random variable $f(X(t))$

provided $X(0) = x$

For the Hamiltonian $H = -\Delta/2 + V(x)$ consider the initial-value problem

$$\frac{du(x,t)}{dt} = \left(\frac{\Delta}{2} - V(x)\right)u(x,t) \qquad (2)$$

with $x \in \mathbf{R}^d$, $d=1$ and $u(x,0)=1$ The Feynman–Kac solution to this equation is

$$u(x,t) = E\exp[-\int_0^t V(x(s))ds], \tag{3}$$

where x(t) is a Brownian motion trajectory and E is the average value of the exponential term with respect to these trajectories and for $V \in K_\nu$, the Kato class of potentials. Numerical work with bare Feynman-Kac procedure employing modern computers was reported [18] for the first time for few electron systems after forty years of the original work [7c] and seemed to be really useful for calculating atomic ground state. A fairly good success in atomic physics motivated us to apply it Condensed matter Physics[19]. Here in this paper we apply Feynman-Kac procedure to Bose-Einstein Condensate of $Rb^{87}$ to visualize the non-equilibrium phenomena connected to the ballistic expansion of a 1d Bose gas.

In terms of Feynman-Kac solution the density function becomes[20]

$$\rho = \left| E\exp[-\int_0^t V(x(s))ds] \right|^2 \tag{9}$$

First we prepare the gas in the ground state of a hard-wall box and let the gas expand to a larger box. Eventually we observe the variation of density with position.

### 1.3. Numerical details and the precision of the calculations.

### 1.3.1 Numerical details:

The formalism given in the Section 1 is valid for any arbitrary dimensions d( for a system of N particles in three dimensions $d = 3N$). Generalizations of the class of potential functions for which Eqns 2 and 3 are valid are given by Simon [21] and include most physically interesting potentials, positive or negative, including, in particular, potentials with $1/x$ singularities. It can be argued that the functions determined by Eq(3) will be the one with the lowest energy of all possible functions independent of symmetry. Although other interpretations are interesting, the simplest is that the Brownian motion distribution is just a useful mathematical construction which allows one to extract the other physically relevant quantities like density, mean square displacement along with the ground and the excited state energy of a quantum mechanical system. In numerical implementation of the Eq(2) the 3N dimensional Brownian motion is replaced by 3N independent, properly scaled one dimensional random walks as follows. For a given time t and integer n and l define [22] the vector in $R^{3N}$.

$$W(l) \equiv W(t,n,l) = (w_1^1(t,n,l), w_2^1(t,n,l), w_3^1(t,n,l), \ldots \ldots w_1^N(t,n,l), w_2^N(t,n,l), w_3^N(t,n,l))$$

where $w_j^i(t,n,l) = \sum_{k=1}^{l} \frac{\epsilon_{jk}^i}{\sqrt{n}}$

with $w_j^i(0,n,l)=0$ for $i = 1,2,\ldots\ldots N; j = 1,2,3 \ and \ l = 1,2,\ldots.nt$. Here $\epsilon$ is chosen independently and randomly with probability $P$ for all $i,j,k$ such that $P = (\epsilon_{jk}^i = 1) = P(\epsilon_{jk}^i = -1) = \frac{1}{2}$. It is known by an invariance principle [23] that for every $\nu$ and $W(l)$ defined above .

$$\lim_{n\to\infty} P(\frac{1}{n}\sum_{l=1}^{nt} V(W(l))) \leq \nu$$

$$=P(\int_0^t V(X(s))ds \leq \nu$$

Consequently for large n,

$$P[\exp(-\frac{1}{n}(\int_0^t V(X(s))ds) \leq v]$$
$$\approx P[\exp(-\frac{1}{n}\sum_{l=1}^{nt} V(W(l))) \leq v]$$

By generating $N_{rep}$ independent replications $Z_1, Z_2, \ldots Z_{N_{rep}}$ of $Z_m = \exp(-\frac{1}{n}\sum_{l=1}^{nt} V(W(l)))$
And using the law of large numbers, $(Z_1 + Z_2 + \cdots + Z_{N_{rep}})/N_{rep} = Z(t)$ is an approximation to Eq(2)

Here $W^m(l), m = 1,2 \ldots N_{rep}$ denotes the $m^{th}$ realization of $W(l)$ out of $N_{rep}$ independently run simulations. In the limit of large t and $N_{rep}$ this approximation approaches an equality and forms the basis of a computational scheme for the solution of a many particle system with a prescribed symmetry.

To show the error analysis, next we consider the simplest case of calculating ground state density of trapped non-interacting BEC in 1d. To achieve that we choose $V_{int} = 0$ at $t = t_0$. Underneath we provide the results for the densities and the error associated it.

### 1.3.2 Precision of the data associated with the calculations of Ground state density of non-interacting BEC in 1d with number of paths($N_{rep}$) = 100000, n=900 and t=1

| x   | Exact solution | FK solution | Exact density | FK density (this work) | error    |
|-----|----------------|-------------|---------------|------------------------|----------|
| 0   | 0.80501        | 0.80508     | 0.64804       | 0.64815                | -0.00011 |
| .1  | 0.80195        | 0.80167     | 0.64312       | 0.64268                | 0.00044  |
| .6  | 0.70189        | 0.70019     | 0.49264       | 0.49027                | 0.00237  |
| 1.1 | 0.50780        | 0.50566     | 0.25786       | 0.25569                | 0.00217  |
| 1.6 | 0.30369        | 0.30200     | 0.09220       | 0.09120                | 0.001    |

For more details in numerical analysis, one should take a look at one of our previous papers [24]
The above data can be represented pictorially as follows:

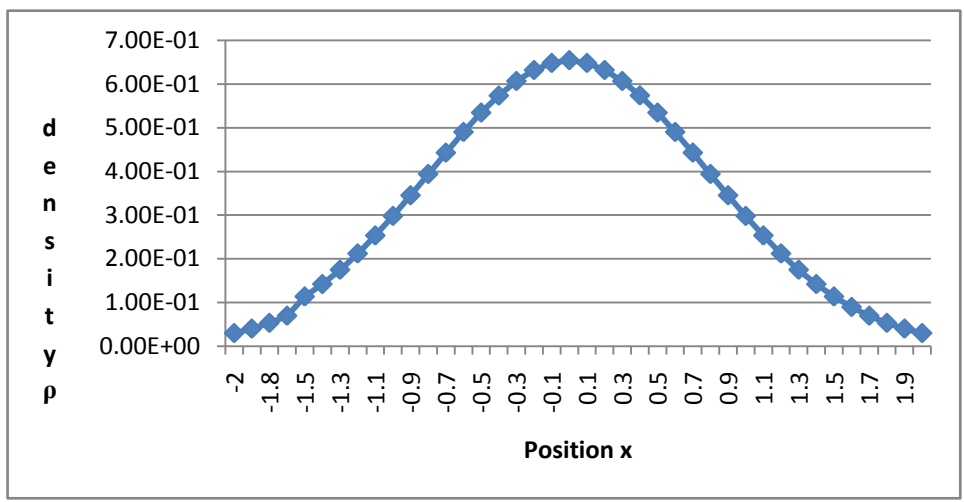

Fig 1 Ground state density of non interacting BEC in 1d in a trap potential

## 2. Results:

Case I. Release from a harmonic trap in the case of weak and strong interaction strength[2]

Table I Values of the scaled[13] coupling strength $g_D^{'}$ for Rb for different trap frequencies $\omega_{||}/2\pi$ and $a_{\perp}^{'} = 0.1$

| $\omega_{||}/2\pi$ | $\omega_{||}$ | $a_0(Rb)(m)$ | $a_{||}(Rb)m$ | $a_0^{'}(Rb) = a_0/a_{||}$ | $g_D^{'}$ |
|---|---|---|---|---|---|
| $10^2$ | 628 | $52 \times 10^{-10}$ | $.1091 \times 10^{-5}$ | $4.7663 \times 10^{-3}$ | 1.02 |
| $10^3$ | 6280 | | $0.0345 \times 10^{-5}$ | $1.506 \times 10^{-2}$ | 3.81 |
| $10^4$ | 62800 | | $0.01091 \times 10^{-5}$ | $4.7663 \times 10^{-2}$ | 28.73 |

We calculate the density of a Bose gas consisting of 100 delta interacting bosons placed infinite potential box of length '$a_{||}$'. Now calculating the densities for three different values of $a_{||}$, we get the following plots for the weak and strong interaction strengths.

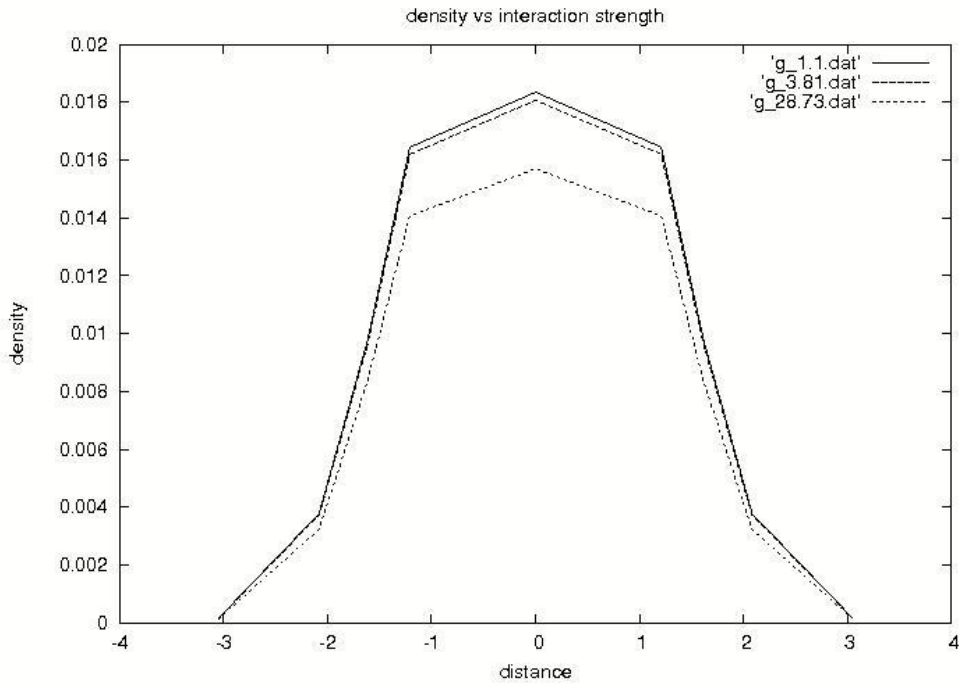

Fig 3 Density vs distance curve for different interaction strength.

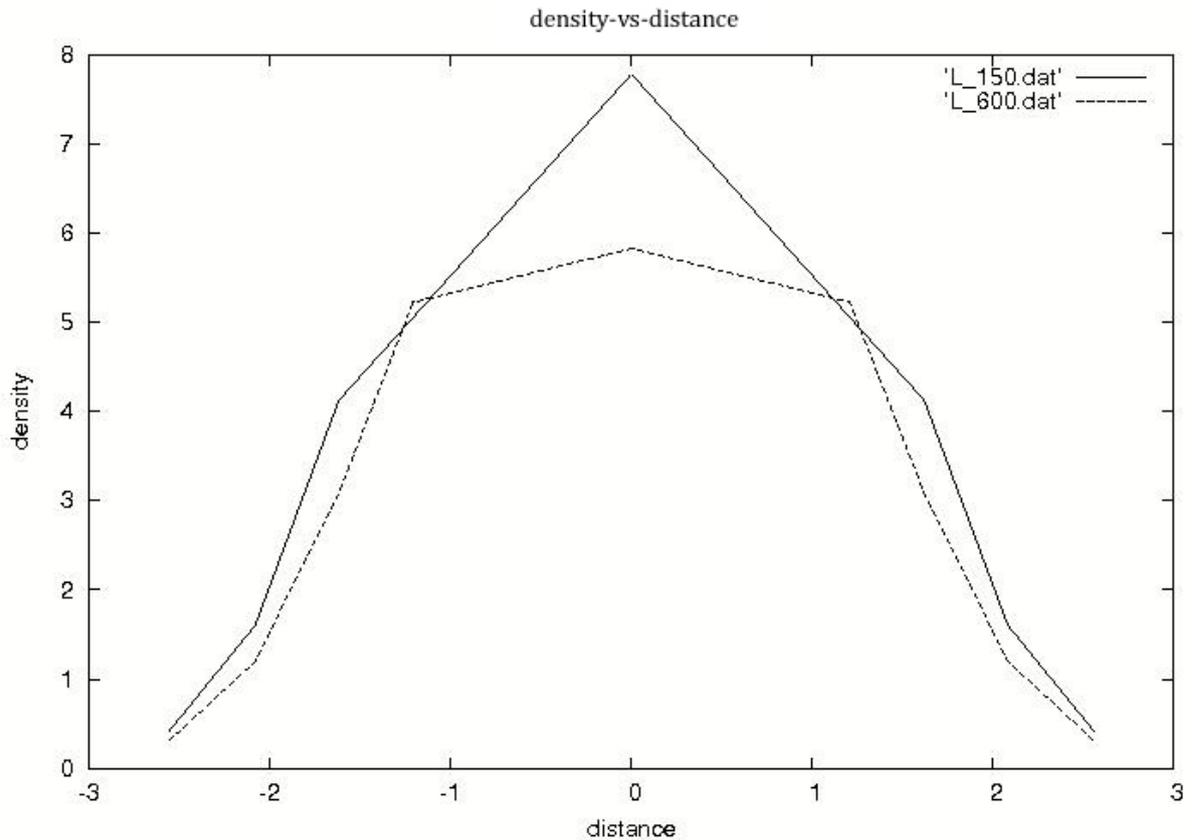

Fig 4   density vs distance curve for box length L=150 and L=600.

### 3. Discussions and Perspectives:

Following the of parametrization in the previous work by Zöller et al[13], we generate the density distributions for different finite values of coupling strength and see that self-similarity is recovered for delta interacting LL gas only for a release from a harmonic trap and for the interaction in the weak and strong limit as shown in Fig 3. In Fig 4, for bosons undergoing a free expansion from an initial hard wall box of size 150 to final length of size L=600,the self similarity and the scale invariance are not observed. One can have a 1D scale invariant expansion in a harmonic trap when the interactions are infinitely strong. For finite coupling constant the expansion is not scale invariant. However in Fig 3 we saw that when bosons are initially in a harmonic trap and the interaction is too weak or too strong the evolution of density is self similar as predicted by the theoretical work of Campbell et al[2] . Similarly, if the trapping potential is a box, the 1D expansion following the sudden switch-off of the trapping potential is never scale invariant, whether the coupling constant is finite or infinite. The only way of enforcing scale invariance for a TG gas released from a trap is modify the trapping potential during the expansion as discussed in del Campo-Boshier [5].

If the interactions are finite, 1D scale invariance emerges only when the coupling constant is modified during the expansion. That would suffice for a harmonic trap. The 1D scale invariant dynamics of a 1D Bose gas with the finite interaction requires not only changing the trapping potential as prescribed by shortcuts during the expansion process but also to modulate the interactions[25]. The relevant numerical simulations are currently underway.

**Acknowledgements:** One of the authors(SD) acknowledges the partial support by the Women Scientist Project Scheme(SR/WOS A/PS-32/2009) of Department of Science and Technology, New Delhi and The Icfai Foundation for Higher Education. The authors wish to thank Dr A del Campo for many useful comments.